# Ho:YAG Thin-Disk Laser with 230 W Multimode and 150 W Single-Mode Output

Xiyi Wang, Xudong Yan, Weichao Yao, and Yuxin Leng

*Abstract*—We report a continuous-wave Ho:YAG thin-disk laser operating at the hundred-watt power level. In multimode operation, the laser delivers a maximum output power of 230 W, with a slope efficiency of 35.9% and an optical-to-optical conversion efficiency of 35.3%. In single-mode operation, an output power of 152.3 W is achieved, with beam quality factors $M^2$ of 1.08 and 1.06 in the horizontal and vertical directions, respectively.

## I. Introduction

Lasers operating in the 2 μm spectral region have been widely used in environmental monitoring, scientific research, and material processing [1]. Motivated by these applications, the 2 μm laser systems have attracted extensive research interest around the world. Solid-state lasers based on $Tm^{3+}$- and $Ho^{3+}$-doped gain media represent a primary route for generating high-power 2 μm laser, with typical materials including $Tm^{3+}$- or $Ho^{3+}$- doped YAG, YAP, YLF, et.al [2-5]. A notable advantage of Tm lasers is their compatibility with direct laser diode pumping at ~800 nm. In this configuration, the quantum efficiency can approach 2 due to the cross-relaxation process. In contrast, Ho lasers exhibit higher conversion efficiency and lower thermal loading at high output power levels with in-band pumping scheme. To date, Ho lasers have achieved output powers on the order of several hundred watts while maintaining high pulse energies [3-5].

However, despite the low thermal loading enabled by in-band pumping in Ho lasers, thermal effects—such as crystal damage, transverse mode degradation, and depolarization in rod-shaped geometries—remain major challenges at high power levels [3]. Thin-disk geometry, which reduces the gain medium thickness to the hundred-micron scale, effectively mitigates thermal loading in the crystal by nearly one-dimensional heat flow. To date, kW level output power with near diffraction-limited beam quality has been demonstrated in Yb:YAG thin-disk lasers [6]. In the 2 μm wavelength region, however, the output power of thin-disk laser remains limited to the ~100 W level [7], with a significant difference compared with Yb:YAG thin-disk lasers and even Ho:YAG rod lasers [4]. The limitation arises mainly from two factors: one is the lack of high-power flat-top Tm fiber laser pump sources; the other is the disk crystal—typically Ho:YAG—cannot be heavily doped, as the conversion efficiency will be restricted by the up-conversion effect. The latter has restricted the minimum disk thickness and limited the tolerable pump intensity.

In this work, we redesigned the pump source to enable the Tm fiber laser to generate a nearly circular flat-top intensity distribution. In addition, the pump power is increased by enlarging the pump spot diameter on the Ho:YAG thin-disk. Using these approaches, the Ho:YAG thin-disk laser achieves record output power of 230 W with an optical-to-optical efficiency of 35% in multimode operation, and 152.3 W in single-mode operation.

## II. Experimental Setup

The experimental setup is illustrated in Fig. 1. A Ho:YAG thin-disk crystal with a doping concentration of 1.6 at.% is used as the gain medium. The disk has a thickness of 210 μm, a diameter of 12 mm, and a radius of curvature (RoC) of 30 m. The front surface of the disk is anti-reflection coated at 1900-2200 nm, while the rear surface is highly reflective in the same wavelength range. The disk is mounted on a diamond heat sink and water-cooled at 16 °C. To improve the pump absorption, a 48-pass pumping scheme is employed, resulting in a total absorption of approximately 87%. A 1 kW 1907 nm Tm fiber laser serves as the pump source. Since Tm fiber lasers typically operate in single or few-mode regimes [8], tens of meters of mode-adapted fiber is used to reshape the transverse intensity of the power-combined Tm fiber laser into a flat-top profile, as shown in Fig. 1(a). The pump beam is then imaged onto the thin-disk with a diameter of 4.5 mm. The maximum pump power is limited to 650 W by the thermal limit of the Ho:YAG thin-disk, corresponding to a pump power density of approximately 4.1 kW/cm$^2$.

For multimode operation, a compact V-shaped cavity is employed, consisting of a concave mirror (RoC = -1 m) and an output coupler, as shown in Fig. 1(a). The calculated Gaussian mode diameter on the thin-disk is 1.8 mm, which is significantly smaller than the pump beam diameter, resulting

This work was supported by the National Natural Science Foundation of China (62505331, 12388102) and sponsored by Natural Science Foundation of Shanghai (25ZR1402532). *(Corresponding author: Weichao Yao).*

Xiyi Wang is with the State Key Laboratory of Ultra-intense Laser Science and Technology, Shanghai Institute of Optics and Fine Mechanics, Chinese Academy of Sciences, Shanghai 201800, China, and with the School of Physical Science and Technology, ShanghaiTech University, Shanghai 201210, China (e-mail: wangxy@siom.ac.cn).

Xudong Yan, Weichao Yao, and Yuxin Leng are with the State Key Laboratory of Ultra-intense Laser Science and Technology, Shanghai Institute of Optics and Fine Mechanics, Chinese Academy of Sciences, Shanghai 201800, China (e-mail: xudong_work_college@outlook.com; wyao@siom.ac.cn; lengyuxin@mail.siom.ac.cn).

in multimode operation. Four output couplers with different transmittances ($T_{OC}$ = 1%, 2%, 3%, and 5%) at 2000-2200 nm are used to optimize the output power.

For single-mode laser experiment, the cavity was extended to improve mode selectivity, comprising a plane mirror, a concave mirror (RoC = -1 m), a concave mirror (RoC = -0.5 m), and an output coupler, as shown in Fig. 1(b). The estimated cavity mode diameter on the thin-disk is 3.9 mm, corresponding to an 87% overlap between the cavity and pump modes.

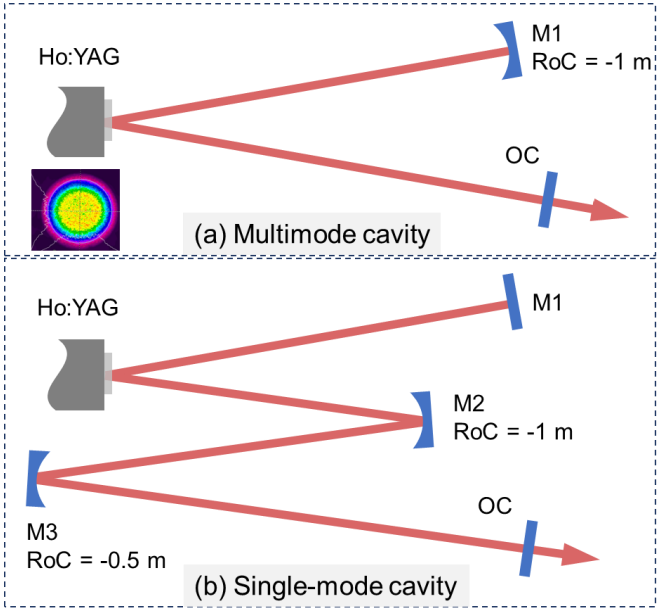

**Fig. 1.** Experimental setup of the Ho:YAG thin-disk laser. (a) Multimode cavity. Inset: beam intensity distribution of the pump source. (b) Single-mode cavity. RoC: radius of curvature; OC: output coupler.

### III. RESULTS AND DISCUSSIONS

Figure 2(a) and Fig. 2(b) present the continuous-wave output power and optical-to-optical conversion efficiency with different output transmittances, respectively. The slope efficiencies for output transmittances of 1%, 2%, 3%, and 5% are 32.8%, 35.6%, 34.6%, and 28.2%, respectively. These values are lower than previously reported results [7], mainly due to pump losses caused by suboptimal reflectivity of optical components in the pump shaping system, which has caused considerable loss of pump power in the multi-pass reflection system.

In our experiment, we have measured the dependence of thin-disk temperature on the pump power. The disk temperature $T$ (in °C) exhibits a linear relationship with the pump power $P$, expressed as $T = 19 + 0.16 \cdot P$. We estimated that when the pump power reaches 650 W (corresponding to a pump power density of 4.1 kW/cm$^2$), the temperature of the Ho:YAG thin-disk will reach the upper limit of ~120 °C for safe operation; therefore, we have set a restriction on the maximum pump power. A maximum output power of 230 W was achieved with a 2% $T_{OC}$ at 650 W pump power, representing the highest output power from a 2 µm thin-disk laser to date. The low output transmittance is attributed to a low gain of the thin-disk. Notably, this low transmittance also leads to an intracavity power exceeding 10 kW, which may induce thermal lensing in cavity optics [7]. Increasing the number of laser reflections on the disk will solve this problem, as this approach can increase the effective crystal length, thus supporting a higher output transmittance. Additionally, the optical-to-optical conversion efficiency exhibits a saturation trend and decline, which may also come from the thermal effects of optical components in the disk module, such as lenses and parabolic mirrors.

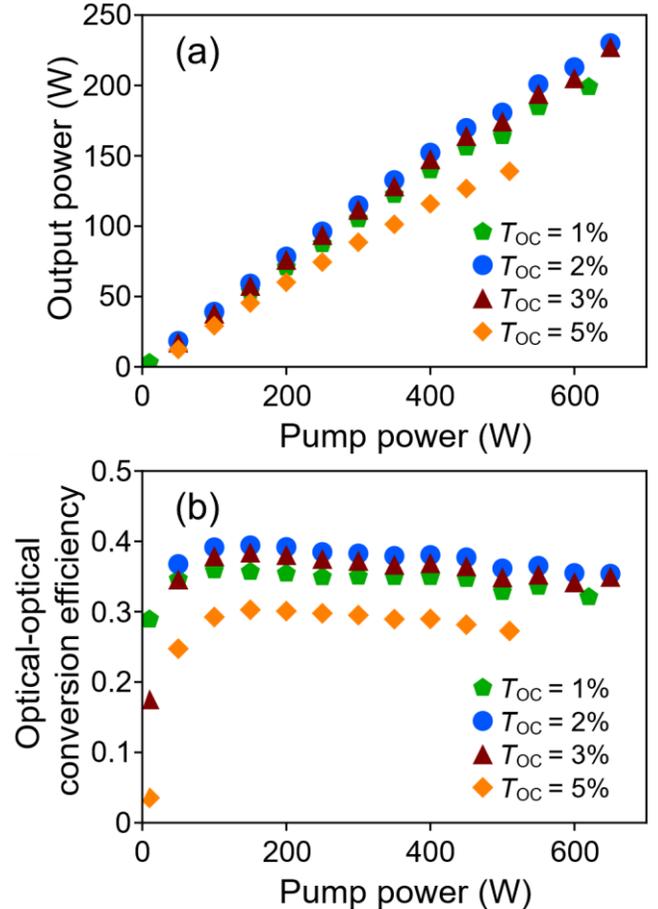

**Fig. 2.** Multimode cavity laser performance. (a) Output power with different output transmittances. (b) Optical-to-optical conversion efficiency versus output transmittance.

Figure 3 presents the laser spectra for different $T_{OC}$ values. The Ho:YAG thin-disk laser exhibits dual-wavelength oscillation at 2090 nm and 2096 nm. As the $T_{OC}$ increases, a blue shift of the laser wavelength is observed. This behavior is attributed to a high population inversion in the Ho:YAG crystal at higher $T_{OC}$, thereby causing the gain peak wavelength to shift toward shorter wavelengths.

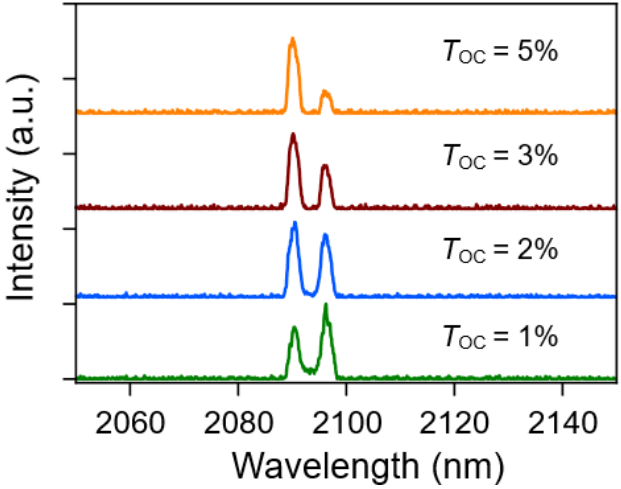

**Fig. 3.** Laser spectra with different output transmittances.

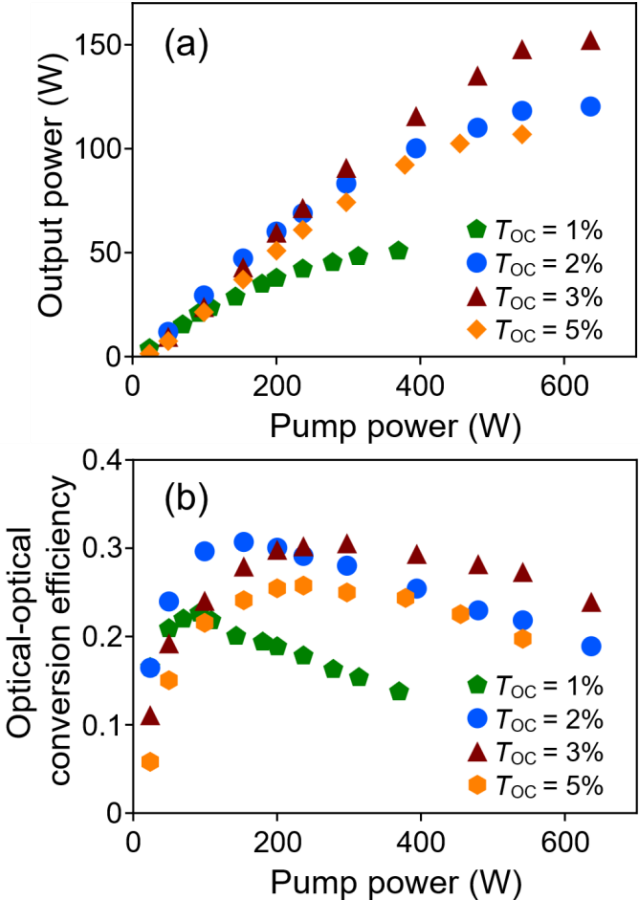

**Fig. 4.** Single-mode cavity laser performance. (a) Output power with different output transmittances. (b) Optical-to-optical efficiency versus output transmittance.

As mentioned earlier regarding the single-mode experimental setup, the estimated cavity mode diameter on the thin-disk is 3.9 mm, and the mode matching ratio is approximately 87%, this has increased the loss of high-order modes and inevitably led to a decrease in conversion efficiency. Figure 4 shows the output power and optical-to-optical conversion efficiency for different $T_{OC}$ values, respectively. At a maximum pump power of 650 W, an output power of 152.3 W was obtained with a 3% $T_{OC}$, corresponding to an optical-to-optical conversion efficiency of 23.9%. Notably, the optimized $T_{OC}$ for single-mode operation is slightly higher than that for multimode operation. This is because the beam is smaller on Ho:YAG disk in single-mode cavity, resulting in a stronger gain saturation, enabling a higher output transmittance. Power saturation at high pump powers may also be attributed to excessively high intracavity power, as the output power level at which saturation occurs exhibits an approximately proportional relationship with the output coupler transmittance.

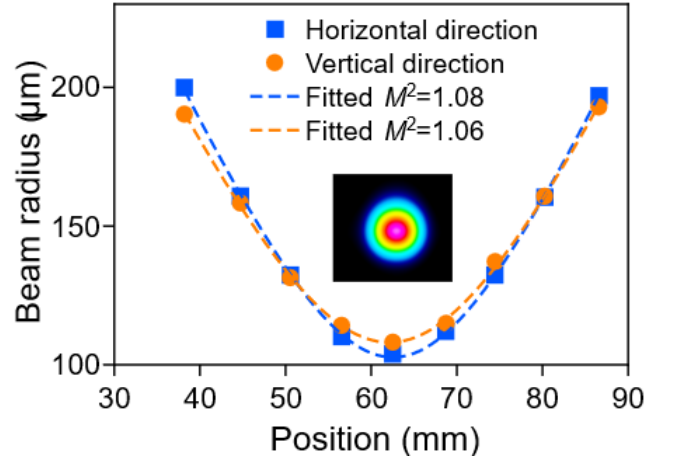

**Fig. 5.** Beam quality $M^2$ measurement results at 115 W output power with 3% $T_{OC}$. Inset: beam intensity profile near the beam waist.

Figure 5 shows the beam quality $M^2$ measurement at an output power of 115 W with a 3% $T_{OC}$, confirming a standard fundamental mode with no obvious astigmatism. Given a large radius of curvature of the Ho:YAG thin-disk crystal, the shift in the resonator operating point induced by thermal effects—which typically cause curvature variation—can be neglected. Consequently, the laser is expected to maintain single-mode operation even at the maximum output power.

For further power scaling, the limitations come from three aspects: pumping loss in the multi-pass pumping module, low absorption efficiency, and low laser gain. Accordingly, several measures should be taken to optimize the Ho:YAG thin-disk system. First, the pump loss can be reduced by redesigning the high-reflection coating of mirrors, which would directly improve the pump efficiency. Second, the absorption can be improved by optimizing the Ho$^{3+}$ doping concentration. However, this optimization requires a careful trade-off between the absorption and the up-conversion effect. For example, a 2 at.% doping concentration will increase the pump absorption to 92% while maintaining a high conversion efficiency [7]. This doping concentration is likely better for Ho:YAG thin-disk geometry. Third, the laser gain can be improved by increasing the number of passes of the laser beam on the disk. This approach not only enhances the effective gain length but also enables the use of a higher output coupler transmittance, thereby reducing the intracavity power and increasing the output power. With these measures, significant power scaling can be expected even in the oscillator configuration, particularly considering that the pump source retains a large power margin, with its maximum available power reaching 1 kW at 1907 nm.

## IV. Conclusion

In conclusion, we have demonstrated a high-power continuous-wave Ho:YAG thin-disk laser with output powers of 230 W in multimode and 152.3 W in single-mode, representing the highest values reported so far from a 2 μm thin-disk laser. We further identify the key limitations of current 2 μm thin-disk laser and outline potential strategies for further optimization. Given the significant development potential of 2 μm thin-disk lasers, Ho:YAG thin-disk laser are expected to play an important role in high-power laser systems, particularly for scaling toward ultrafast regimes.